\newif\ifColor\Colortrue \Colorfalse
\Colortrue
\PassOptionsToPackage{super,comma,numbers,sort&compress}{natbib}  
\documentclass[preprint, 5p, 8pt, lefttitle]{elsarticle}

\usepackage{xpatch}
\xpatchcmd{\MaketitleBox}{\hrule\vskip12pt}{\vspace{-2\baselineskip}}{}{}
\xpatchcmd{\MaketitleBox}{\hrule}{}{}{}

\usepackage[switch, left]{lineno}
\modulolinenumbers[1]
\usepackage[utf8]{inputenc}
\usepackage{amssymb}
\usepackage{bm}
\usepackage{amsmath}
\usepackage{amsfonts}
\usepackage{upgreek}
\usepackage{gensymb}
\usepackage{widetext}
\usepackage{setspace}
\usepackage{booktabs}
\usepackage[dvipsnames]{xcolor}
\usepackage[english]{babel}
\usepackage{mathtools}
\usepackage{mhchem}
\usepackage{hyperref}
\usepackage[final]{microtype}
\usepackage{csquotes,lmodern}
\usepackage{subcaption}
\usepackage{etoolbox}
\usepackage{color}
\usepackage{multibib}
\newcites{methods}{References}
\usepackage[labelfont=bf]{caption}

\usepackage{sidecap}
\usepackage{adjustbox}
\newcommand{\PFunit}[1]{{$\,$mWm$^{-1}$K$^{-2}$}}
\newcommand{\kunit}[1]{{$\,$Wm$^{-1}$K$^{-1}$}}
\newcommand{\Sunit}[1]{{$\,\upmu$VK$^{-1}$}}
\newcommand{\EF}[1]{{$E_\text{F}$}}
\newcommand{\Eg}[1]{{$E_\text{g}$}}
\newcommand{\Seeband}{{\textsf{\textit{SeeBand}}}}

\makeatletter
\renewenvironment{abstract}{\global\setbox\absbox=\vbox\bgroup
  \hsize=\textwidth\def\baselinestretch{0}%
 \par\unskip\noindent\unskip\ignorespaces}
 {\egroup}

\def\keyword{%
  \def\sep{\unskip, }%
 \def\MSC{\@ifnextchar[{\@MSC}{\@MSC[2000]}}
  \def\@MSC[##1]{\par\leavevmode\hbox {\it ##1~MSC:\space}}%
 \def\PACS{\par\leavevmode\hbox {\it PACS:\space}}%
  \def\JEL{\par\leavevmode\hbox {\it JEL:\space}}%
 \global\setbox\keybox=\vbox\bgroup\hsize=\textwidth
  \normalsize\normalfont\def\baselinestretch{0}
 \parskip\z@
  \noindent\textit{Some important words: }   <--- Edit as necessary
  \raggedright                         
  \ignorespaces}

\def\ps@pprintTitle{%
     \let\@oddhead\@empty
     \let\@evenhead\@empty
     \def\@oddfoot{\footnotesize\itshape
      \ifx\@journal\@empty  
       \else\@journal\fi\hfill\today}%
     \let\@evenfoot\@oddfoot}
     
\makeatother

\begin{document}
\begin{frontmatter}

\title{\doublespacing\textbf{\fontsize{22}{22}\selectfont{\textsf{\textit{SeeBand}: A highly efficient, interactive tool for analyzing electronic transport data}}}}

\author[1]{\textsf{\textbf{Michael Parzer}}\corref{cor1}}
\author[1]{\textsf{\textbf{Alexander Riss}}
\corref{cor1}}
\author[1]{\textsf{\textbf{Fabian Garmroudi}}
\corref{cor1}}
\author[2,3]{\textsf{\textbf{Johannes de Boor}}}
\author[4,5]{\textsf{\textbf{Takao Mori}}}
\author[1]{\textsf{\textbf{Ernst Bauer}}}
\cortext[cor1]{e-mail: michael.parzer@tuwien.ac.at, alexander.riss@tuwien.ac.at, f.garmroudi@gmx.at}
\address[1]{\textsf{Institute of Solid State Physics, TU Wien, 1040 Vienna, Austria}}
\address[2]{\textsf{ Institute of Materials Research, German Aerospace Center (DLR), D–51147 Cologne, Germany}}
\address[3]{\textsf{University of Duisburg-Essen, Faculty of Engineering, Institute of Technology for Nanostructures
(NST) and CENIDE, 47057 Duisburg, Germany}}
\address[4]{\textsf{International Center for Materials Nanoarchitectonics (WPI-MANA), National Institute for Materials Science, Tsukuba, Japan}}
\address[5]{\textsf{Graduate School of Pure and Applied Sciences, University of Tsukuba, Tsukuba, Japan}}
\selectlanguage{english}
\journal{submitted to npj Computational Materials}

\begin{abstract}
\noindent \textsf{\fontsize{10}{10}\selectfont{Linking the fundamental physics of band structure and scattering theory with macroscopic features such as measurable bulk thermoelectric transport properties is indispensable for a thorough understanding of transport phenomena and ensures more targeted and methodical experimental research.
Here, we introduce \Seeband{}, a highly efficient and interactive fitting tool based on Boltzmann transport theory. A fully integrated user interface including a visualization tool enables real-time comparison and connection between the electronic band structure (EBS) and microscopic transport properties. It allows simultaneous analysis of data for the Seebeck coefficient $S$, resistivity $\rho$ and Hall coefficient $R_\text{H}$ to identify suitable EBS models and extract the underlying microscopic material parameters and additional information from the model. Crucially, the EBS can be obtained by directly fitting the temperature-dependent properties of a single sample, which goes beyond previous approaches that look into doping dependencies.  Finally, the combination of neural-network-assisted initial guesses and an efficient subsequent fitting routine allows for a rapid processing of big datasets, facilitating high-throughput analyses to identify underlying, yet undiscovered dependencies, thereby guiding material design.}}
\end{abstract}

\end{frontmatter}


\section*{Introduction}
\noindent Rapidly increasing numbers of experimental data call for enhanced tools of data analysis. In the field of thermoelectrics, tens of thousands of temperature-dependent transport data have been accumulated. The \textsf{\textit{Starrydata2}} open web database \cite{katsura2019data} alone aggregates TE property data from 51,985 samples and 8,956 papers as of May 2024. The analysis of these immense amounts of data was often forsaken or performed within a simple single-parabolic-band model in the unipolar regime with a single dominant scattering mechanism in the original publications. Since most high-performance TE materials are, however, often far more nuanced, \textit{e.g.}, showcasing bipolar transport \cite{bahk2014enhancing,zhang2015suppressing,witting2019thermoelectric}, convergence of multiple electronic bands \cite{pei2011convergence,pei2012band,tan2015codoping,shi2024global} or complex carrier scattering \cite{shuai2017tuning,ren2020establishing,garmroudi2023high}, the lack of adequate and accurate analyses leads to limited or even erroneous interpretations and researchers reverting to time-consuming empirical studies. 

Tuning and manipulating electronic transport is crucial for realizing next-generation technologies, such as quantum computing or energy harvesting, which are expected to play an integral role in our future society \cite{priya2009energy,pecunia2023roadmap,ladd2010quantum,huang2020superconducting}. In this context, it is imperative to understand the behavior of charge carriers to enable informed and targeted research and accelerate material's development. This is of utter importance across various branches of the physical sciences, encompassing, for instance, the pursuit of high-temperature superconductivity in fundamental condensed matter physics, or the optimization of TE transport in materials science. By comparing experimental data with theoretical predictions, profound insights into electronic correlations, microstructure, and structure-property relationships can be obtained. Furthermore, the synergy between transport data analysis and theoretical calculations validates respective results, establishing a robust foundation for understanding material properties. While novel theoretical frameworks and codes capable of directly calculating transport properties from \textit{ab initio} electronic structure calculations are constantly being developed \cite{madsen2006boltztrap,madsen2018boltztrap2,
cepellotti2022phoebe,pickem2023linretrace,graziosi2023electra}, there is currently no common tool that can derive microscopic EBS parameters in reverse from experimental transport data. Instead, various groups have developed similar frameworks for data analysis, demonstrating the demand for such measurement data analysis, which complements theoretical calculations \citep{lee1980electron,bulusu2008review,schmidt2014using,kamila2020non,naithani2022developing,agrawal2024dopant}. The inaccessibility of these tools, however, has led to the common usage of much simpler formalisms for material characterization like the \textit{weighted mobility}, primarily derived from approximations to the single-parabolic-band model \citep{may2009characterization, kim2015characterization, gibbs2015band, snyder2020weighted}. 

Here, we introduce \Seeband, a software package for electronic transport data analysis, based on the Boltzmann transport formalism and parabolic band approximation. \textsf{\textit{SeeBand}} constitutes a highly efficient fitting tool that allows visualizing and linking the interdependence of fundamental, microscopic features such as the EBS around the Fermi energy $E_\text{F}$ with macroscopic quantities such as measured temperature dependencies of electronic transport. Our code features a unique, neural-network-assisted least-squares fitting algorithm, which is able to concurrently handle the temperature-dependent electrical conductivity, Seebeck coefficient and Hall coefficient, different but complementary electronic transport properties that are most ubiquitously and easily measured in a laboratory. The highly efficient numerical framework of \Seeband{} allows extremely rapid processing of data, which can be leveraged in the high-throughput analyses of big datasets, as we demonstrate in this paper on the basis of around 1000 datasets of half-Heusler compounds.

While \Seeband{} is designed to enhance the comprehension of TE transport in narrow-gap semiconductors, fostering more targeted and efficient experimental research, its utility extends beyond thermoelectrics to other domains within materials science and solid-state physics, where a profound understanding of charge transport, especially in semimetallic and semiconducting systems, is crucial.

This paper is organized as follows: First, we provide a brief overview of the theoretical foundations underlying charge transport in TE materials, along with their implementation in the \Seeband{} code. In the subsequent section, a practical illustration and demonstration of usage (see also the Supplemental Video as an additional user guide) is offered. Following this, section 3 delves into real-world test examples, highlighting the applicability across different material classes. Finally, we underscore that the high efficiency and capability for automated fitting in the \Seeband{} code enables high-throughput data analysis, exemplified through the examination of 1000 TE property datasets of half-Heusler compounds within the \textsf{\textit{StarryData2}} database \cite{katsura2019data}. 

\section{Theoretical background}
\label{sec:theor}
\begin{figure}[t!]
\centering
\includegraphics[width=0.5\textwidth]{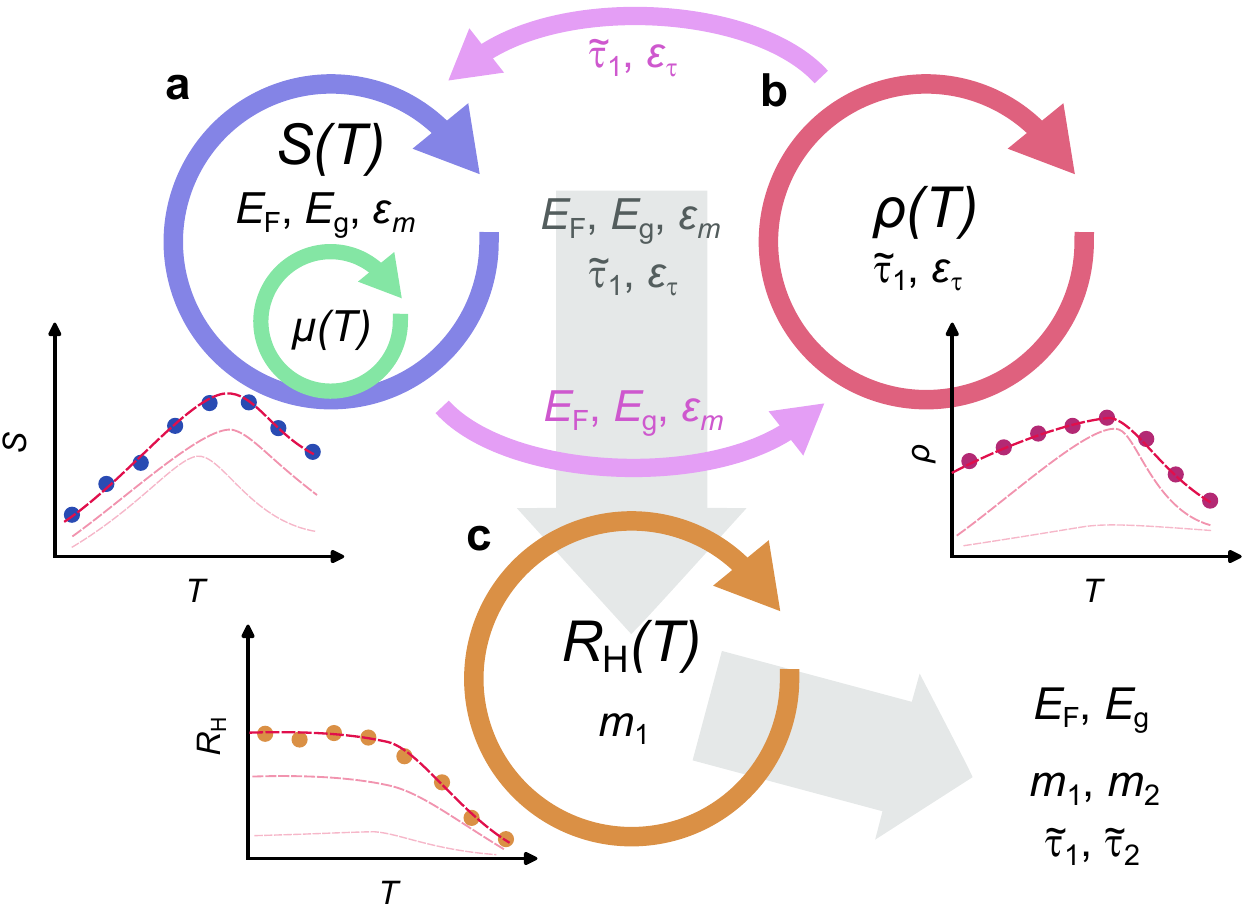}
\caption{\textbf{$\vert$\,\Seeband 's fitting algorithm}. \textbf{a} Modeling the temperature-dependent Seebeck coefficient $S(T)$ with two parabolic bands requires three independent fitting parameters, \textit{i.\,e.}, the Fermi energy $E_\text{F}$, the band gap $E_\text{g}$ and a weighting parameter $\varepsilon_m=(N_1m_2)/(N_2m_1)$, which depends on the band degeneracies and effective masses. During the fitting routine, the chemical potential $\mu(T)$ is calculated each time a band parameter is changed. \textbf{b} The band parameters obtained from the Seebeck fit are directly used for modeling the temperature-dependent resistivity $\rho(T)$, for which two additional parameters have to be taken into consideration, namely the prefactors of the electron relaxation time of the first band $\widetilde{\tau}_{1}$ and the ratio of the relaxation time prefactors of the two bands $\varepsilon_\tau=\widetilde{\tau}_2/\widetilde{\tau}_1$. These parameters are transferred between the two fit loops and iteratively optimized. \textbf{c} In a final single-parameter fit of the temperature-dependent Hall coefficient $R_\text{H}$(T), the absolute value of the effective masses can be determined.}
\label{fig:Fig1}
\end{figure}
\noindent
The \Seeband{} code is constructed on a framework based on the Boltzmann transport theory and the parabolic band approximation.
In the semiclassical Boltzmann theory, the generalized transport coefficients $L_\alpha(\mu,T)$, which depends on both temperature and the chemical potential $\mu(T)$. $L_\alpha(\mu,T)$ can be written in terms of a general transport distribution function $\Sigma(E)$:
\begin{equation}
\label{BTT}
L_\alpha (\mu,T)=q^2\int_{-\infty}^{\infty} \Sigma(E) (E-\mu)^\alpha\left(-\frac{\partial f}{\partial E}\right) \, dE\;.
\end{equation} 
$\Sigma(E)$ depends on the EBS and dominant scattering processes:
\begin{equation}
\label{BTT_Sigma}
\Sigma(E)=\sum_{i=1}^{N} \frac{1}{(2\pi)^3}\int \tau_{i,k} v_{i,k} v_{i,k} \delta(E-E_{i,k}) \, d^3k\;.
\end{equation}
The summation is performed over $N$ bands and integration is done over the first Brillouin zone, with $v_{i,k}$ and $\tau_{i,k}$ being the band velocities and electron relaxation times, respectively. 
For parabolic bands in 3D materials, the spherical symmetry of the Fermi surface in $k$ space allows simplification of \autoref{BTT_Sigma} to
\begin{equation}
\Sigma(E)=D(E)v^2(E) \tau(E)\;,
\end{equation}
with $D(E)$ being the density of states. Since $D(E)\propto E^{1/2}$, $v^2(E)\propto E$ and $\tau(E)\propto E^{-1/2}$ for acoustic-phonon and alloy-disorder scattering, it follows that $\Sigma(E)\propto E$. Thus, the number of parameters is reduced and \autoref{BTT} can be written in terms of Fermi integrals for a single band
\begin{equation}
F_j(\eta(T))=\int_0^\infty \frac{\xi^j}{e^{\xi-\eta}+1} \, d\xi\;.
\end{equation}
Here, $\xi=E/(k_\text{B}T)$ and $\eta=\mu/(k_\text{B}T)$ represent the reduced energy and chemical potential, respectively. Utilizing Fermi integrals significantly enhances the efficiency of the fitting process, as robust solutions for these integrals are readily available. Since these integrals are at the heart of the \Seeband{} framework and enter all the expressions of the transport equations, we especially focused on optimizing the numerical efficiency of their computation (for details see SI). This enables a rapid processing of the data, where the Fermi integrals can be evaluated up to $10^6$ times per second.
\subsection{Equations used for the fitting process}
\noindent Below, we briefly list the equations and fit parameters (highlighted in bold) which are used in the refinement process. For a single parabolic band, the equations for the transport properties are given by \citep{may2012introduction}
\begin{equation}
S(T)=\frac{k_\text{B}}{e}\left[\bm{\eta}-\frac{2F_1(\bm{\eta)}}{F_0(\bm{\eta})}\right]\;,
\end{equation}
\begin{equation}
\rho(T)=\left[\frac{2\sqrt{2}e^2\,k_\text{B}T}{\pi^2\hbar^3}\bm{\left(\frac{\widetilde{\tau{}}}{m}\right)}T^\gamma F_0\left(\eta\right)\right]^{-1}\;,
\end{equation}
\begin{equation}
R_\text{H}(T)=\frac{\pi^2\hbar^3}{2e\left(2\,\bm{m}\,k_\text{B}T\right)^{3/2}}\frac{F_{-1/2}\left(\eta\right)}{F_{0}^2\left(\eta\right)}\;.
\end{equation}
Here, $k_\text{B}$, $e$ and $\hbar$ are physical constants, Boltzmann constant, electron charge and reduced Planckian constant, respectively, while $\eta$, $\widetilde{\tau}$ and $m$ are the reduced chemical potential, the scattering prefactor and the effective mass, respectively, representing the refinement parameters.
For two parabolic bands (2PB), the single-band contributions are weighted and summed up appropriately \cite{putley1975galvano,may2012introduction} (a general derivation of the transport fitting equations for $N$ bands is given in the SI), yielding for the Seebeck coefficient, the resistivity and the Hall coefficient. \\\\

\begin{widetext}
\begin{equation}
\label{Seebeck2PB}
S_\text{2PB}(T)=\frac{F_0(\bm{\eta})S_1(\bm{\eta},T)+\varepsilon_\tau/\bm{\varepsilon_m}\,F_0(\bm{\eta}-\bm{E_\textbf{g}})S_2(\bm{\eta}-\bm{E_\textbf{g}},T)}{F_0(\bm{\eta})+\varepsilon_\tau/\bm{\varepsilon_m}\,F_0(\bm{\eta}-\bm{E_\textbf{g}})}\;,
\end{equation}
\begin{equation}
\label{Rho2PB}
\rho_\text{2PB}(T)=\left[\frac{2\sqrt{2}e^2\,k_\text{B}T^{\gamma+1}N_1 }{\pi^2\hbar^3\,m_1}\bm{\widetilde{\tau}_{1}}\left(F_0(\eta) + \bm{\varepsilon_\tau}/\varepsilon_m F_0(\eta-E_\textbf{g})\right) \right]^{-1}\;,
\end{equation}
\begin{equation}
\label{RH_2PB}
R_\text{H,2PB}(T)=\frac{3 \pi^2\hbar^3}{2e\left(2k_\text{B}T\right)^{3/2}}\bm{m_1}^{-3/2}\left[\frac{F_{-1/2}(\eta)+\left(N_1/N_2\right)^{5/2}\varepsilon_\tau^2/\varepsilon_m^{7/2}\,F_{-1/2}(\eta-E_\textbf{g})}{\left(F_0(\eta)+\varepsilon_\tau/\varepsilon_mF_0(\eta-E_\textbf{g})\right)^2}\right]\;.
\end{equation}\\

\end{widetext}

\noindent The scope of \Seeband{} is to extract relevant microscopic EBS parameters from experiments, by fitting data using above equations.
Fig.\,\ref{fig:Fig1} sketches the refinement technique implemented in the \Seeband{} software. In the case of two parabolic bands, modelling the Seebeck coefficient via \autoref{Seebeck2PB} requires three independent fitting parameters: the position of the Fermi level $\bm{E_\textbf{F}}$, which determines the temperature-dependent chemical potential $\bm{\eta}$, the band gap $\bm{E_\textbf{g}}$, \textit{i.\,e.} the energy difference between the two band edges, as well as a weighting parameter $\bm{\varepsilon_\tau/\varepsilon_m}$. Here, $\bm{\varepsilon_\tau}=\widetilde{\tau}_{1}/\widetilde{\tau}_{2}$ denotes the ratio of the electron relaxation time prefactors of the two bands and $\bm{\varepsilon_m}=(N_1m_2)/(N_2m_1)$ is the band mass ratio $m_2/m_1$, weighted with the band degeneracies $N_1$ and $N_2$. Note that during the fitting procedure, the chemical potential has to be continuously recalculated in each step when the EBS parameters $\bm{E_\textbf{F}}$, $\bm{E_\textbf{g}}$ and $\bm{\varepsilon_m}$ are changed (see Fig.\,\ref{fig:Fig1}a). Most importantly, each of these parameters has a unique effect on the temperature-dependent behavior of $S(T)$, enabling robust and unambiguous conclusions to be drawn when fitting experimental data of $S(T)$.

Other transport properties like the \sloppy temperature-dependent resistivity $\rho(T)$ and Hall coefficient $R_\text{H}(T)$ should in principle be describable by the very same EBS model. To further enhance the robustness of the obtained fit parameters, simultaneous analyses of all these temperature-dependent transport properties ($S(T)$, $\rho(T)$, $R_\text{H}(T)$) can be performed in the \Seeband\, framework.

Since $\rho(T)$ depends on the relaxation times $\tau_i$ of the charge carriers associated with their respective electronic bands, additional fit parameters have to be introduced. In the case of two parabolic bands, the minimal number of additional free parameters is given by (i) the electron relaxation time $\bm{\widetilde{\tau}_1}$ of the first band and (ii) the ratio of the electronic relaxation times between the two bands $\bm{\varepsilon_\tau}=\widetilde{\tau}_1/\widetilde{\tau}_2$. For parabolic bands and hence spherical Fermi surfaces, $\tau$ can be written in a general form \citep{parzer2024extending}

\begin{equation}
\tau(E,T)=\frac{\widetilde{\tau}\,T^\gamma} {D(E)\,m}=\widetilde{\tau}\,T^\gamma E^{-1/2}m^{-3/2}\,.
\end{equation}

\noindent Here, $\gamma$ is a scattering-specific parameter and $\widetilde{\tau}$ denotes an energy- and temperature-independent prefactor. The most important charge carrier scattering mechanisms in thermoelectric materials are (i) acoustic phonon and (ii) alloy disorder scattering, for which $\gamma=-1$ and 0, respectively. In the metallic limit, this yields the well-known linear resistivity behavior expected for electron-phonon scattering $\rho(T)\propto\tau^{-1}\propto T$ at elevated temperatures and a temperature-independent residual resistivity $\rho_0$ for alloy-disorder scattering. The electron relaxation time is further determined by material-specific parameters \cite{bardeen1950deformation,
krishnamurthy1985generalized,may2012introduction}, which can be subsumed in a constant prefactor

\begin{equation}
\widetilde{\tau}_\text{ph}=\frac{\pi\hbar^4 v_\text{l}^2\rho_\text{m}}{\sqrt{2}k_\text{B}\Xi_\text{ph}^2}\,,
\end{equation}

\begin{equation}
\widetilde{\tau}_\text{dis}=\frac{\sqrt{2}\pi\hbar^4 n_A}{x_A(1-x_A)U_\text{dis}^2}\,.
\end{equation}

\noindent For dominant acoustic phonon scattering, the relevant physical parameters defining $\widetilde{\tau}$ are the electron-phonon interaction, which may be represented by an acoustic deformation potential $\Xi_\text{ph}$, the longitudinal sound velocity $v_\text{l}$ and the material density $\rho_\text{m}$. In the case of alloy-disorder scattering, $\widetilde{\tau}$ depends on the atomic fraction of alloyed atoms $x_A$, their particle density $n_A$ and on a scattering potential $U_\text{dis}$, which accounts for the random fluctuations of the periodic lattice potential caused by the random substitution of alloyed atoms. In the \Seeband{}, framework, the user can decide whether one or both of these scattering processes are taken into account. When both scattering processes are considered at the same time, Mathiessen's rule is employed to calculate the total relaxation time $\tau_\text{tot}=\tau_\text{ph}\tau_\text{dis}/(\tau_\text{ph}+\tau_\text{dis})$.

\begin{figure*}[t!]
\centering
\includegraphics[width=0.9\textwidth]{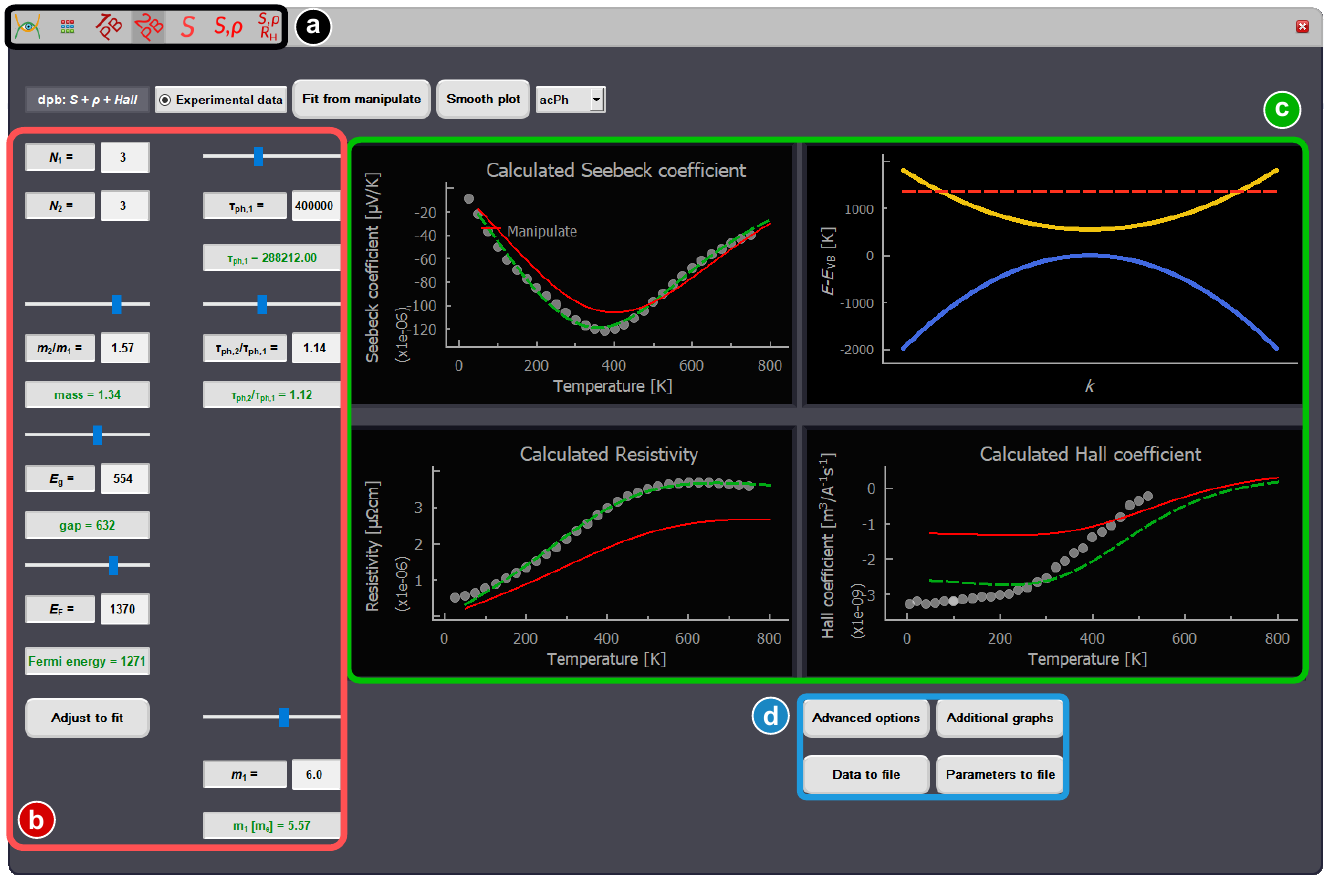}
\caption{\textbf{$\vert$\,User interface of \Seeband .} Exemplary snapshot of the user interface of the \Seeband{} software. \textbf{a} Toolbar allowing to toggle between different windows. The user can choose between a single- or two-parabolic band analysis of the transport properties. Additionally, the interface offers to either analyze just the Seebeck coefficient, or a combination of different transport properties simultaneously, depending on what data are available. \textbf{b} Different fit and input parameters, as explained in \autoref{sec:theor} and \autoref{sec:demuse}. \textbf{c} Imported experimental data and partially adjusted effective electronic structure in the graphical user interface (GUI). The experimental data are depicted as gray circles, while the calculated transport properties corresponding to the adjustable parameters of the sliders are depicted as the solid red lines. The panel on the top right shows the effective band structure close to the Fermi level $E_\text{F}$ (dashed red line), corresponding to the transport properties displayed as a red solid lines. After pressing the \textit{Fit from manipulate} button, the mathematically best fit of band parameters for the respective data and the resulting transport coefficients are displayed in green. \textbf{d} Buttons for further features including additional graphs, like the Seebeck coefficients and conductivities of the individual bands, the temperature dependence of the chemical potential, as well as the electronic thermal conductivity including bipolar contributions.}
\label{fig:Fig2}
\end{figure*}

As sketched in Fig.\,\ref{fig:Fig1}\,a,b the \Seeband\, code uses an iterative loop, wherein the parameters obtained from the Seebeck modelling are directly transferred to evaluate $\rho(T)$. 
Note that for the refinement of $S(T)$ only the mass ratio 1/$\varepsilon_m$ is adjusted, while for the refinement of $\rho(T)$, the ratio of the relaxation time prefactors $\varepsilon_\tau$ itself is the fitting parameter. Both steps are reiterated until convergence is achieved. Finally, when data of the temperature-dependent Hall coefficient are available, the absolute value of the effective masses can be evaluated from the band mass ratio $\varepsilon_m$ by performing a final single-parameter ($\bm{m_1}$) fit of $R_\text{H}(T)$ (see Fig.\,\ref{fig:Fig1}c). This procedure allows for a minimal number of free parameters to be used, ensuring more robust and unambiguous results. Simultaneously, the user is able to extract a variety of relevant physical parameters concerning the EBS and scattering times directly from synergistic temperature-dependent measurements on a single sample.
\section{Demonstration of use}
\label{sec:demuse}
\noindent To enhance the user experience and aid understanding of how various band parameters influence transport properties, we developed a user-friendly graphical interface (GUI) for \Seeband\, providing interactive control over the parameters. As the user modifies the values, the transport properties update in real-time, along with a visualization of the corresponding effective band structure. Additionally, the GUI facilitates direct control over the fitting process and allows extraction of further information from the results. Data can also be exported for external analysis. A concise overview of the GUI is provided here, with a more comprehensive user guide available SI2 of the supplementary material.
\begin{figure*}[t!]
\centering
\includegraphics[width=0.95\textwidth]{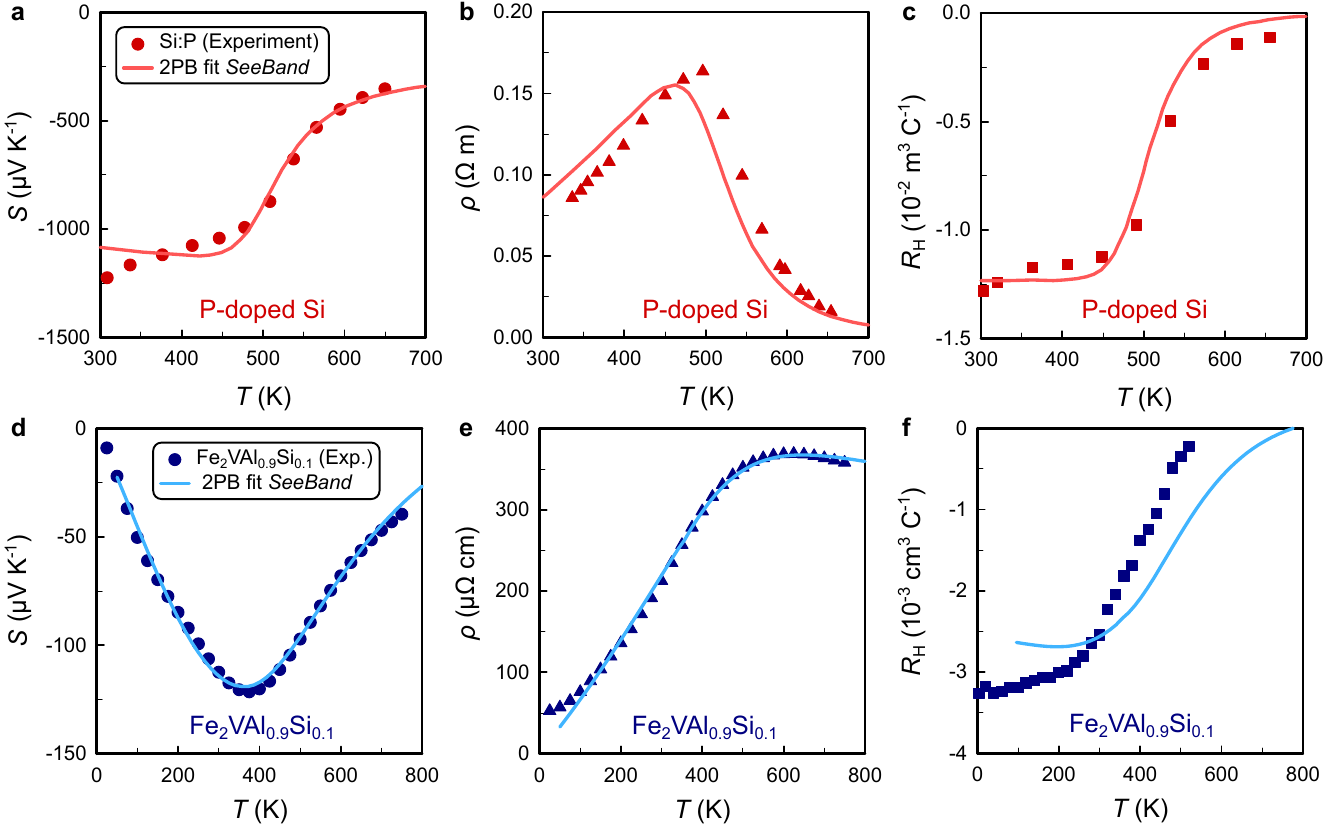}
\caption{\textbf{$\vert$\,Two test examples, a wide- and a narrow-gap semiconductor system, for accurate electronic transport analysis with \Seeband .} Temperature-dependent Seebeck coefficient, electrical resistivity and Hall coefficient of, \textbf{a\,--\,c}, phosphorus-doped silicon \cite{agrawal2024multi} and, \textbf{d\,--\,f}, full-Heusler-type \ce{Fe2VAl_{0.9}Si_{0.1}} \cite{garmroudi2021boosting}. Symbols represent experimental data and solid lines least-squares fits from the simultaneous analyses of all transport properties as sketched in \autoref{fig:Fig1}.}
\label{fig:Fig3}
\end{figure*}

\subsection{Graphical user interface}
The user interface for the two-parabolic-band model and all three transport coefficients is illustrated in \autoref{fig:Fig2}.
The top toolbar (a) of the GUI provides access to different windows to input data, adjust and fit the transport properties based on available data and switch between the single- and two-parabolic-band model.

Relevant band parameters can be entered in the text fields and sliders on the left (b). After successfully fitting the data, the results of the fitting are displayed in green. The four panels on the right (c) show the three transport coefficients ($S$, $\rho$, $R_\text{H}$) along with the effective band structure derived from the band parameters. Gray circles represent measurement data, while the red and green lines show the transport properties derived from the input parameters and the least-squares fit, respectively.

The \textit{Adjust to fit} button sets the user controlled parameters (and the resulting calculated transport properties) to the values of the prior fitting, enabling the analysis of how modifications of those parameters impact the transport properties. The buttons at the bottom (d) allow the user to set limits for the fitting process, access additional information such as electronic thermal conductivity or the temperature dependence of the chemical potential $\mu(T)$, and export relevant data to well-arranged text files.

\subsection{Workflow}
The workflow proceeds as follows:
First, the measurement data are input using the input window (the first symbol on the toolbar). Next, the appropriate model is selected, based on the measurement data and the desired fitting model. The user has the option of only modelling $S(T)$ (see Fig.\,\ref{fig:Fig1}a), a combined iterative analysis of the Seebeck coefficient and resistivity (see Fig.\,\ref{fig:Fig1}a,b) or, when Hall data are also available, a combined analysis of $S(T)$, $\rho(T)$ and $R_\text{H}(T)$ together. Additionally, the user can choose between a single and two-parabolic band model. The trained neural network (NN) will provide an initial guess for the relevant band parameters, usually offering a sufficiently good starting point for the fit to converge (see SI1 of the supplemental material for more information on the neural network). Before starting the fit, the relevant scattering mechanisms are chosen, and limits for all fitting parameters can be adjusted, if necessary. Finally, the fitting process can be initiated to determine the optimal electronic structure that best describes the measurement data within the chosen model. Subsequent to the fit, the interface allows for further analysis by manually adjusting parameters while observing the changes in $S$, $\rho$, and $R_\text{H}$. Moreover, additional information can be displayed based on the number of bands and provided measurement data.

\section{Test examples}
\noindent In this section, to elucidate the usefulness and validity of the refinement tools, we present practical use-cases of well-studied materials, where reliable comparison of our modeling results with the electronic structure can be made. To demonstrate applicability across different material classes with distinct electronic structures, we study P-doped silicon (Si:P) as a wide-gap semiconductor, and the full-Heusler compound \ce{Fe2VAl_{0.9}Si_{0.1}} as a narrow-gap thermoelectric material. Comprehensive temperature-dependent transport data for the Seebeck coefficient, the electrical resistivity and the Hall coefficient are available in the literature for both materials \cite{agrawal2024multi,garmroudi2021boosting}.

\subsection{Phosphorous-doped silicon}

Fig.\,\ref{fig:Fig3}\,a-c show the temperature-dependent Seebeck coefficient $S(T)$, electrical resistivity $\rho(T)$ as well as the Hall coefficient $R_\text{H}(T)$ of Si:P \cite{agrawal2024multi} in the temperature range 300\,--\,650\,K. The temperature-dependent transport properties were modeled assuming acoustic-phonon scattering using the multi-stage refinement algorithm implemented in the \Seeband{} code (\textit{cf.} Fig.\,\ref{fig:Fig1}) and the least-squares fits are depicted as red solid lines in Fig.\,\ref{fig:Fig3}a-c. Very good agreement between experimental data and the theoretical curves is apparent for all three independent transport properties, highlighting the robustness of the underlying theory. 

The obtained EBS parameters from the least-squares fits are summarized in \autoref{tab:Si:P} and compared to values from various literature studies, yielding very good overall agreement. This demonstrates the ability of combined analyses of various transport properties for extracting robust information of the EBS.  

In lightly doped semiconductors like Si:P, the Fermi energy is situated at the localized electronic states arising from the dopant's impurity levels, which lie about 0.044 eV below the conduction band edge \cite{long1959hall}. In \Seeband , the localized nature of such localized in-gap impurity states cannot be captured, since only parabolic bands, \textit{i.\,e.} free carriers, are assumed. Nonetheless, the obtained position of $E_\text{F}$, which lies only about $0.02$\,meV within the conduction band, aligns with the notion of $E_\text{F}$ being placed at the conduction band edge.

The band gap, extracted with \Seeband , yields $E_\text{g}=1.08\,$eV, in near-perfect agreement with the real band gap of silicon, which is 1.12\,eV at room temperature and decreases down to $\approx 1.03$\,eV at 650\,K \cite{bludau1974temperature,green1990intrinsic}.\\ 
The conduction band of Si forms six equivalent ellipsoidal constant energy surfaces, associated with different values of the effective mass in longitudinal and transverse directions. The valence band complex consists of a doubly degenerate light and heavy hole band plus a \textit{split-off} hole band \cite{green1990intrinsic}. In our transport modeling, these bands are assumed parabolic and isotropic and the derived values of the effective mass have to be considered an effective average. Despite these simplifications, the derived values of $\varepsilon_m$ and $m_1$ are in good agreement with those from literature \cite{green1990intrinsic}.

\subsection{Full-Heusler \ce{Fe2VAl_{0.9}Si_{0.1}}}
Most state-of-the-art thermoelectric materials are narrow-gap semiconductors. To demonstrate the applicability of \Seeband{} to such systems, we studied thermoelectric transport in \ce{Fe2VAl}-based full-Heusler compounds, which have emerged as an important class of thermoelectric materials for potential applications in the temperature range 300\,--\,500\,K \cite{hinterleitner2019thermoelectric,
mikami2021power,garmroudi2021boosting,
fukuta2022improving,bourgault2023unlocking}. In \ce{Fe2VAl}, a threefold degenerate valence and conduction band form an almost zero-gap electronic structure \cite{anand2020thermoelectric,hinterleitner2021electronic, garmroudi2022large}. Depending on the doping concentration, bipolar conduction sets in already below room temperature, requiring a two-band framework for properly describing the electronic transport properties. Devising suitable EBS models in such materials is crucial, since it allows, for example, estimating the bipolar contribution to the thermal conductivity, which severely limits the performance and $zT$ and cannot be differentiated experimentally by making use of the Wiedemann-Franz law. Neglecting the pivotal role of minority carrier contributions on the transport properties or the chemical potential leads to erroneous interpretations and predictions of the maximum achievable figure of merit \citep{kuo2020systematic}. 

Fig.\,\ref{fig:Fig3}d-f shows experimental data of $S(T)$, $\rho(T)$ and $R_\text{H}(T)$ together with least-squares fits in a 2PB model framework for $n$-doped \ce{Fe2VAl_{0.9}Si_{0.1}}. The onset of bipolar conduction manifests itself in a distinct maximum of $S(T)$ at $T\approx 400\,$K, and also $\rho(T)$ at slightly higher temperatures, as well as a kink in $R_\text{H}(T)$ at around 200\,--\,300\,K.

Theoretical curves yield excellent alignment not only with the experimental $S(T)$ and $\rho(T)$, but also for the Hall coefficient $R_\text{H}(T)$, considering that only a single free fitting parameter has been used for the latter. The obtained EBS parameters are once again compared to previous information existing in the literature (see \autoref{tab:Si:P}), for which we find very good quantitative agreement. Crucially, while previous modeling attempts are very time-consuming or require the synthesis of samples with different doping concentrations, \Seeband{} enables a highly efficient treatment of temperature-dependent data of a single sample. The whole multi-stage refinement process usually takes only a couple of seconds ($\approx 1.5$\,s for \ce{Fe2VAl_{0.9}Si_{0.1}}), which makes \Seeband{} a potent tool for high-throughput analyses of large data sets.

\begin{table}
\centering 
\textbf{Phosphorus-doped Si}
\vspace{2mm}
\begin{tabular}{|c|c|c|} 
\hline 
\text{Parameter} & \text{Literature} & \text{Refinement} \\ \hline
$E_\text{F}$ & 1.076\,eV\cite{long1959hall} & 1.08\,eV\,+\,0.02\,meV \\ \hline
$E_\text{g}$ & 1.03\,--\,1.12\,eV\cite{bludau1974temperature,green1990intrinsic} & 1.08\,eV \\ \hline
$\varepsilon_m$ & 0.6\cite{green1990intrinsic} & 0.74  \\ \hline
$\widetilde{\tau}_{1}$ & - & $3.44\cdot10^{5}\,$kg$^2$\,m\,K \\ \hline
$\varepsilon_\tau$ & - & 1 \\ \hline
$m_1$ & 0.55\,$m_\text{e}$\cite{green1990intrinsic} & 0.49\,$m_\text{e}$ \\ \hline
$m_2$ & 0.33\,$m_\text{e}$\cite{green1990intrinsic} & 0.36\,$m_\text{e}$ \\ \hline
\end{tabular}
\\
\centering 
\textbf{Heusler-type \ce{Fe2VAl_{0.9}Si_{0.1}}}
\begin{tabular}{|c|c|c|} 
\hline 
\text{Parameter} & \text{Literature} & \text{Refinement} \\ \hline
$E_\text{F}$ & - & 0.110\,eV \\ \hline
$E_\text{g}$ & 0.03\,--\,0.1\,eV\cite{nishino1997semiconductorlike,okamura2000pseudogap,anand2020thermoelectric} & 0.055\,eV \\ \hline
$\varepsilon_m$& 2.71\cite{anand2020thermoelectric} & 1.34  \\ \hline
$\widetilde{\tau}_{1}$ & - & $3.00\cdot10^{5}\,$kg$^2$\,m\,K\,\,\,\,\,\\ \hline
$\varepsilon_\tau$ & - & 0.96 \\ \hline
$m_1$ & 2.3\,$m_\text{e}$\cite{anand2020thermoelectric} & 5.6\,$m_\text{e}$ \\ \hline
$m_2$ & 6.1\,$m_\text{e}$\cite{anand2020thermoelectric} & 7.5\,$m_\text{e}$ \\ \hline
\end{tabular}
\caption{\textbf{Comparison of band structure parameters for two test examples.} Microscopic EBS parameters of phosphorus-doped silicon and Heusler-type \ce{Fe2VAl_{0.9}Si_{0.1}}, extracted from simultaneous least-squares fits of temperature-dependent transport data shown in \autoref{fig:Fig3}, utilizing \Seeband . The obtained band parameters are compared to existing values in the literature \cite{long1959hall,bludau1974temperature,
green1990intrinsic,nishino1997semiconductorlike,
okamura2000pseudogap,anand2020thermoelectric}, yielding good agreement. All energies are given with respect to the valence band edge ($E_i-E_\text{VBM}$).}
\label{tab:Si:P}
\end{table}

\begin{figure*}[t!]
\centering
\includegraphics[width=0.95\textwidth]{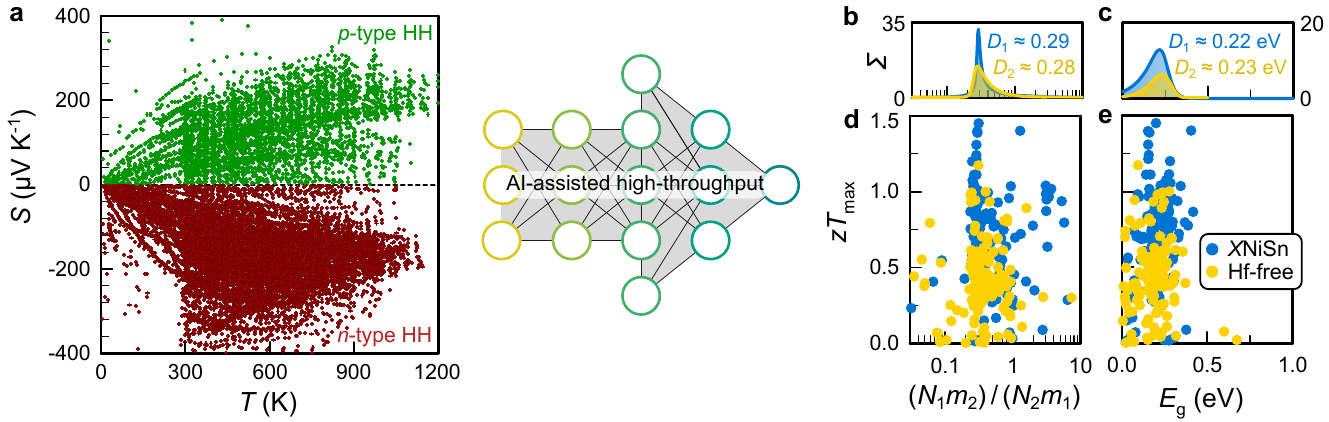}
\caption{\textbf{$\vert$\,Neural-network-assisted high-throughput analysis of hundreds of half-Heusler compounds.} \textbf{a} Temperature-dependent Seebeck coefficient $S(T)$ for hundreds of \textit{p}- and \textit{n}-type half-Heusler compounds from the \textsf{\textit{Starrydata2}} open web database \cite{katsura2019data}. Neural-network-assisted initial guesses and a fast and automated subsequent fitting routine enable high-throughput modelling of $S(T)$. \textbf{b},\,\textbf{c} The statistical frequency $\Sigma$ and modes $D_1(\varepsilon_m,E_\text{g})$ and $D_2(\varepsilon_m,E_\text{g})$ reveal an almost identical effective mass ratio $\varepsilon_m=(N_1m_2)/(N_2m_1)=0.28$ - $0.29$ between conduction and valence bands, as well as an almost identical band gap $E_\text{g}=0.22$ - $0.23\,$eV for Hf-free and Hf-containing $X$NiSn. This confirms that the enhancement of $zT$ for Hf-containing half-Heuslers primarily arises from a reduction of the lattice thermal conductivity and that Hf has a negligible effect on the electronic structure. \textbf{d},\,\textbf{e} Maximum figure of merit $zT_\text{max}$ versus electronic band structure parameters for $X$NiSn half-Heuslers ($X=\text{Ti}$, Zr, Hf), with and without hafnium.}
\label{fig:Fig4}
\end{figure*}

\section{High-throughput data analysis}
\noindent The sheer volume of experimental TE property data available in the literature is immense, with much of it remaining unanalyzed and underutilized. The advent of comprehensive materials databases such as \textsf{\textit{StarryData2}} and \textsf{\textit{MaterialsProject}}, which are progressively incorporating experimental data, underscores the need for advanced tools to handle large datasets. These databases enable the extraction of myriad data, making automated fitting essential. \Seeband{} addresses this need by offering automated fitting capabilities, though it still requires robust start parameters to navigate local minima and avoid divergences from the correct results. To overcome this issue, we trained a neural network (for details see SI) to provide initial parameter guesses, which can be leveraged to facilitate the high-throughput analysis of thermoelectric transport properties across thousands of compounds.

In this section, we demonstrate this on the example of \ce{$X$NiSn}-based half-Heusler compounds, where $X=\text{Ti}$, Zr, Hf. Owing to their exceptional mechanical properties and high $zT$, $n$-type \ce{$X$NiSn} half-Heuslers represent one of the hottest candidates for realizing thermoelectric applications at elevated temperatures \cite{fey2023isa,pecunia2023roadmap,li2024half}. To this aim, we downloaded hundreds of transport datasets of different classes of half-Heusler compounds (see Fig.\,\ref{fig:Fig4}a) and fitted the temperature-dependent Seebeck coefficient employing the AI-assisted framework of \Seeband{}. 

In the high-throughput analysis of these data, we focus on \ce{$X$NiSn}-based half-Heuslers, comparing samples with and without Hf. Fig.\,\ref{fig:Fig4}c,d shows $zT_\text{max}$ as a function of the extracted EBS parameters such as the weighted effective mass ratio $\varepsilon_m$ and the band gap $E_\text{g}$, both of which can be directly derived from fitting $S(T)$ (\textit{cf.} Fig.\,\ref{fig:Fig1}a). Only for less than 5\,\% of the datasets the NN guess was insufficiently accurate for the fit to converge. These datasets were subsequently discarded from the analysis. Despite substantial scatter of the obtained EBS parameters, most likely arising from varying sample and data quality, there exists a clear accumulation, as can be seen from the smoothed statistical frequencies plotted in Fig.\,\ref{fig:Fig4}b,c. The corresponding modes $D_{1,2}(\varepsilon_m,E_\text{g})$ reveal that both Hf-free and Hf-containig \ce{$X$NiSn} half-Heuslers can be desribed with almost identical EBS parameters. This ascertains that the main enhancement of the figure of merit $zT_\text{max}$ stems from a reduction of the lattice thermal conductivity and not from optimizations of the EBS. 

The mode for the band gap is $D_1(E_\text{g})\approx D_2(E_\text{g})=0.22-0.23$\,eV. This value is in very good agreement with what has been reported in the literature up until now. For instance, a simple estimation of the thermal band gap via the Goldschmid-Sharp relation $E_\text{g}=2e\,|S_\text{max}|T_\text{max}$ yields 0.27\,eV \cite{schmitt2015resolving}. For ZrNiSn, Aliev \textit{et al.} derived band gaps of 0.25\,eV using infrared optical spectroscopy and 0.18\,eV from Arrhenius-type activation of $\rho(T)$, while Schmitt \textit{et al.} derived smaller values of around 0.13\,eV, using diffuse optical reflectance measurements \cite{schmitt2015resolving}. It should be noted that the value of $E_\text{g}$ obtained experimentally is significantly smaller than that obtained from DFT calculations, since $E_\text{g}$ in real samples does not correspond to the energy gap between valence and conduction bands but to the gap between intrinsic in-gap impurity bands and the conduction band \cite{ougut1995band,do2014electronic,
zeier2016engineering}. 
The pivotal role of these impurity bands is also reflected in the effective mass ratio $\varepsilon_m=(N_1m_2)/(N_2m_1)=0.28$ - $0.29$. By taking into account the band degeneracies \cite{do2014electronic} and the effective mass of the conduction band carriers $m_2\approx 2.9\,m_\text{e}$ \cite{zeier2016engineering}, we derive an effective hole mass of $m_1\approx 30.5\,m_\text{e}$, which implies that hole-type charge carriers, associated with the impurity bands behave as massive charge carriers, in agreement with the flat dispersions expected from such impurity bands \cite{do2014electronic}. 

\section*{Conclusion}
\noindent Summarizing, we developed a highly efficient, easy-to-use yet powerful refinement tool, \Seeband{}, capable of deriving important physical band structure parameters and relevant information regarding charge carrier scattering, directly from temperature-dependent transport data. The Boltzmann transport theory and parabolic band framework make \Seeband\, especially suitable for the study of thermoelectric materials, which are typically doped semiconductors, where the conducting Fermi surface pockets can often be described by a parabolic band dispersion. \Seeband{} possesses the ability to concurrently analyze temperature-dependent Seebeck, resistivity, and Hall effect data, which guarantees robust and clear results, as demonstrated with two well-studied material classes: P-doped silicon and \ce{Fe2VAl}-based full-Heuslers. Moreover, a trained neural network embedded in the \Seeband{} framework provides initial parameter guesses for the fitting procedure, enabling extremely fast processing of the data and consequently high-throughput data analyses, which we exemplified for about 1000 TE property data sets of half-Heusler compounds. Adequately and efficiently mode1ling electronic charge transport is crucial for rationally designing functional electronic materials such as thermoelectrics, especially considering the vast quantity of data which has been accumulated over the recent decades of research. \Seeband{} in its current form already constitutes a powerful tool for transport data analysis, its potential, however, is immense. By addressing next steps such as expanding the range of selectable scattering mechanisms, incorporating non-parabolic band models, and including formulas for phononic heat transport, \Seeband{} can evolve into a comprehensive tool for analyzing measurement data across various fields of solid-state physics.

\section*{Data availability}
The datasets used and/or analysed during the current study available from the corresponding authors on reasonable request.
\section*{Code availability}
The underlying code and training/validation datasets for this study is available from the Git repository for \Seeband \citep{parzer2024SeeBand_git} and can be accessed via this link [\url{https://github.com/xAngelswordx/SeeBand}]. Also an executable version of the software \Seeband{} is available as a GitHub release in the same repository via [\url{https://github.com/xAngelswordx/SeeBand/releases/tag/Executable}].

\section*{Acknowledgements}
The research in this paper was supported by the Japan Science and Technology Agency (JST) programs MIRAI, No. JPMJMI19A1. The authors thank Nikolas Reumann for fruitful discussions regarding the theoretical framework of \Seeband{}.

\section*{Author contributions}
M.P. and A.R. developed the software \Seeband{}. F.G., M.P. and A.R. and developed the theoretical framework behind the parabolic-band transport model. J. d.B. substantially contributed to both the development of the theoretical framework and the software. A.R. and M.P. developed the neuronal network, implemented in the software for initial parameters guesses. M.P., A.R. and F.G. devised and performed the high-throughput analysis. J. d.B. and E.B. supervised the work. T.M. organized the funding.
F.G., M.P. and A.R. wrote the initial draft. All authors discussed the results and edited the final manuscript.

\section*{Competing interests}
The authors declare they have no competing interests.

\end{document}